\documentclass[aps,twocolumn,groupedaddress,prl]{revtex4}

\usepackage{graphicx}
\usepackage{revsymb}
\begin{document}


\title{AFM local oxidation nanolithography of graphene}

\author{Lishan Weng}
\author{Liyuan Zhang}
\author{Yong P. Chen}
\author{L.~P. Rokhinson}
\affiliation{Birck Nanotechnology Center and Department of Physics, Purdue
University, West Lafayette, Indiana 47907 USA}
\date{\today}

\begin{abstract}
We demonstrate the local oxidation nanopatterning of graphene films by an atomic
force microscope. The technique provides a method to form insulating trenches
in graphene flakes and to fabricate nanodevices with sub-nm precision. We
demonstrate fabrication of a 25-nm-wide nanoribbon and sub-micron size nanoring
from a graphene flake. We also found that we can write either trenches or bumps
on the graphene surface depending on the lithography conditions. We attribute
the bumps to partial oxidation of the surface and incorporation of oxygen into
the graphene lattice.
\end{abstract}


\maketitle

Recently graphene has received special attention due to its remarkable
electronic properties
\cite{Novoselov04,zhang05,Novoselov05,Berger06,Novoselov07,Geim07}. The most
commonly used method for the fabrication of graphene nanodevices has used
conventional electron-beam lithography and subsequent plasma etching
\cite{Russo08,Han07,Chen07,Bunch05,Stampfer08}. On the other hand, alternative
lithography techniques, especially those based on scanning probe microscopy,
have shown great potentials for patterning various materials at nanoscale
\cite{tseng05}. Atomic force microscopy (AFM) based local anodic oxidation
(LAO) lithography has been used to fabricate micro- and nano- structures on
metallic or semiconductor surfaces \cite{snow94,held97,rokhinson02}. In
particular, AFM has been used to cut carbon nanotubes \cite{kim03} or etch
holes in highly oriented pyrolytic graphite (HOPG) \cite{park07}. The
advantages of LAO include the ability to pattern surfaces with nanometer
resolution and to examine devices during the lithography process, and easy
tuning of the fabrication. LAO nanolithography is performed in the ambient
environment and eliminates several fabrication steps, such as photoresist
processing needed in conventional lithography. The main disadvantage of LAO -
shallow oxidation of materials - should not be an issue when pattering a few
layers of graphene. Moreover, atomic resolution of freshly cleaved graphite is
routinely achievable in mainstream scanning probe microscopy, thus atomic
control of oxidation is possible. In this letter, we report direct LAO of
graphene flakes.  As an example, we fabricate a 25-nm-wide nanoribbon and a
sub-micron nanoring in a single layer graphene. We also report that under
certain conditions we can form bumps on the surface of graphene flakes instead
of trenches. This may indicate partial oxidation and incorporation of oxygen
into the graphene lattice instead of formation of volatile carbon oxide.

\begin{figure}[t]
\def\ffile{fig1}
\includegraphics[scale=0.75]{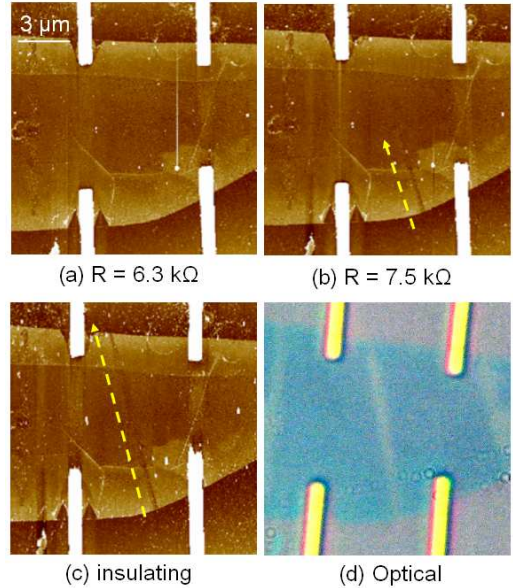}
\caption{(a) AFM image of an uncut graphene flake (thickness $\sim$ 5nm). The
two-terminal resistance (R) from the left to the right end of the flake was
measured to be 6.3 k$\Omega$. The four white bars in the picture are the metal
contacts (whose heights are beyond the limit of the color scale). (b) A trench
was cut from the edge to the middle of the flake, along the direction indicated
with the green dashed arrow. R = 7.5 k$\Omega$. (c) The trench was cut through,
electrically insulating the left and right parts of the flake. (d) Optical
microscope image of the completely cut-through trench.}
\label{\ffile}
\end{figure}

We have characterized LAO patterning of grapheneflakes with thickness
ranging from one to several atomic layers. Our graphene sheets were mechanically
exfoliated from a natural graphite and transferred onto a 300 nm-thick SiO$_2$
on a heavily doped Si substrate. The graphene flakes are identified by their
color contrast under an optical microscope followed by thickness measurements by AFM.
Cr/Au (3 nm/50 nm) electrodes, used for electrical characterization as
discussed below, were fabricated by electron beam lithography and metal
deposition before or after LAO lithography.

We use a Veeco Dimension 3100 AFM system with an environmental enclosure with
controlled humidity. The system has a noise floor $\sim$0.3 nm in the lateral
directions, precluding atomic resolution for graphene. For both imaging and
lithography, a conductive silicon tip was used in a non-contact (tapping) mode
in which constant height is maintained using optical feedback. The sample
substrate is grounded. A small negative bias voltage (amplitude 15-30 V) is
applied on the tip, creating an electric field large enough to induce
electrochemical oxidation on the sample at room temperature.  The bias voltage
is modulated between zero and the set value with a 100-Hz square wave to help
stabilize a water meniscus around the tip. We did not find that electrically
grounding or floating the graphene itself makes significant differences for the
LAO.

We first demonstrate that LAO can be used to electrically isolate different
regions in a graphene flake. Fig.~\ref{fig1}(a) shows a test flake before patterning, with
a resistance, measured from the left to the right end, of 6.3 k$\Omega$.
A line written across half of the flake using LAO results in a small increase of that
resistance to 7.5 k$\Omega$ (Fig.~\ref{fig1}(b)). As the line is
continuously cut through the whole flake in the subsequent lithography step,
the resistance across the line becomes infinite (Fig.~\ref{fig1}(c)). The line is barely seen in the AFM
image but is clearly seen under the optical microscope (Fig.~\ref{fig1}d). The
width of the line from the optical image is overestimated by a factor of 5.

\begin{figure}[t]
\def\ffile{Picture2_0001}
\includegraphics[scale=0.7]{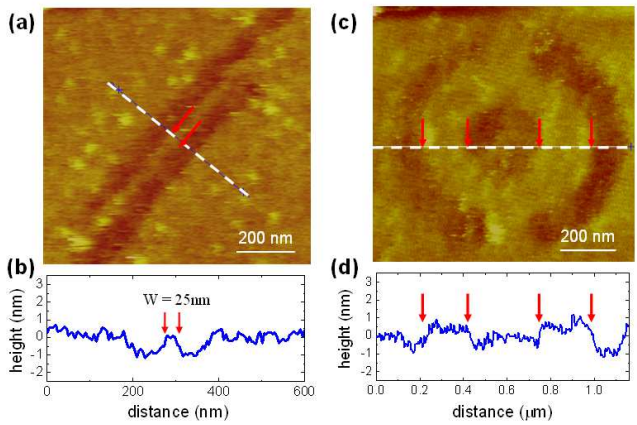}
\caption{ (a) AFM image of a nanoribbon fabricated on a graphene flake with
thickness $\sim$1 nm. The width and length of the ribbon is 25 nm and 800 nm
respectively. (b) Height profile along the dashed line in (a), crossing the
trenches. (c) A nano ring (inner radius $\sim$ 160 nm, outer radius $\sim$380
nm) patterened on a similar graphene flake as used in (a). Note that two long
trenches, not shown in the picture, were subsequently drawn from the
circumference of the ring outward to the edges of the flake to electrically
isolate the ring device. (d) Cross-sectional height profile along the dashed
line in (c).}
\label{\ffile}
\end{figure}

\begin{figure}[t]
\def\ffile{Picture3_0001}
\includegraphics[scale=0.6]{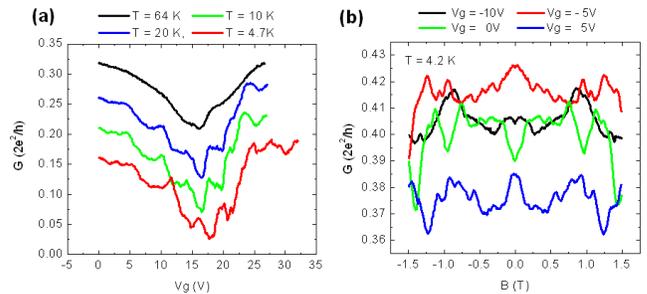}
\caption{(a) Conductance versus gate voltages for several temperatures. Note
that the three curves for $T$ = 20 K, 10 K and 4.7 K were offset by -0.05,
-0.1, and -0.15 (in units of $2e^2/h$) respectively. (b) the conductance versus
magnetic field at various gate voltages measured at 4.2 K.} \label{\ffile}
\end{figure}

Several graphene nanodevices have been fabricated using the LAO technique. In
Fig~\ref{Picture2_0001}a we show a graphene nanoribbon formed between two
LAO-patterned trenches. The width of a 800 nm long ribbon is $\sim$ 25 nm, the
trench depth is equal to the flake thickness of $\sim$1 nm (corresponding to
1-2 layers of graphene). In Fig~\ref{Picture2_0001}c, a ring pattern fabricated
by LAO is shown. The inner and outer radii of the ring are 160 nm and 380 nm
respectively. The width of the conducting region of the ring is 220 nm. We have
characterized the conductance ($G$) of the ring (additional trenches were cut
to confine the current in the flake to flow through the ring). At temperature
($T$) below $\sim$50 K, we observe reproducible fluctuations of conductance as
a function of magnetic field ($B$) or gate voltage ($V_g$), shown in
Fig.~\ref{Picture3_0001}. We attribute these fluctuations to universal
conductance fluctuations (UCF) \cite{staley08,graf07,moriki07,Russo08} in such
a mesoscopic device. From the UCF in $G(B)$ we estimate a phase coherence
length via $B_c\l_{\varphi}^2\approx \phi_0$, where $B_c$ is the half with of
the autocorrelation function of $G(B)$ (after subtracting a smooth background
in $G(B)$) and $\phi_0=h/e$ is a flux quantum. At 4.2 K, thus calculated
$l_{\varphi}\sim 90$ nm is much smaller than the circumference (1.7 $\mu$m) of
the ring, consistent with the absence of Aharonov-Bohm oscillations
\cite{Russo08,recher07} in our device.

\begin{figure*}[t]
\def\ffile{Picture4_0001}
\includegraphics[scale=0.85]{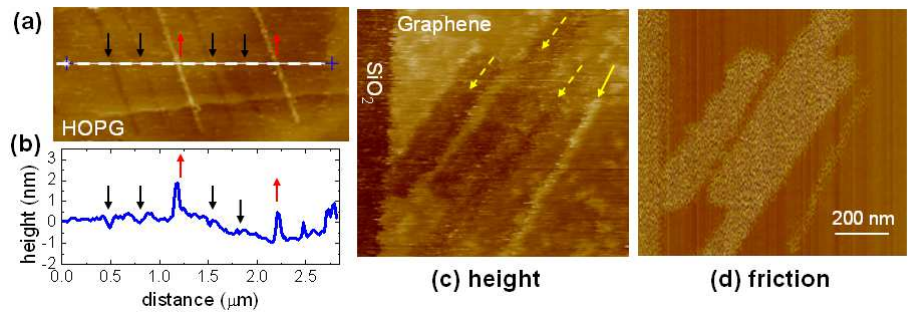}
\caption{AFM images of line patterns created by the LAO technique. (a) Trenches
or bumps were formed on HOPG surface. 6 lines (from left to right) are written
with the same tip bias voltage of $-20$ V while the set point ($SP$, see text
for more details) was cycled through 0.3, 0.2, 0.1, 0.3, 0.2, 0.1 volts,
corresponding to 15\%, 10\%, 5\%, 15\%, 10\% and 5\% of the free-oscillation
amplitude for an unloaded tip. (b) Height profile across the line marked in
(a). Red arrows indicate bumps and black arrows indicate trenches. (c) Three
trenches and one bump were patterned on a graphene flake (with $\sim$1 nm in
thickness). The dashed arrows indicate trenches; the solid arrow indicated a
bump. (d) The frictional force image corresponding to height image (c).}
\label{\ffile}
\end{figure*}

The formation of trenches as described above can be understood as due to the
oxidation of graphene into volatile carbon oxides under the AFM tip
\cite{kondo95}. On the other hand, we have found that for certain LAO
conditions we can form bumps rather than trenches on graphene flakes as well as
on HOPG. The conditions that result in trenches or bumps can be controlled by
the combination of a set point voltage (which controls the tip-sample distance
in the dynamical mode) and tip bias voltage. In general, lower bias voltages
and/or lower set points ($SP$) result in bumps, while higher bias voltages
and/or higher set points result in trenches. In Fig.~\ref{Picture4_0001}a, 6
lines (from left to right) are written with the same tip voltage of $-20$ V
while $SP$ was cycled through 0.3, 0.2, 0.1, 0.3, 0.2, 0.1 volts, which
correspond to 15\%, 10\%, 5\%, 15\%, 10\% and 5\% of the free-oscillation
amplitude for an unloaded tip respectively. While higher $SP=0.3$ and 0.2 V
consistently results in formation of trenches, bumps are written for $SP=0.1$
V. Alternatively, the type of writing can be controlled by the tip bias
voltage. In Fig.~\ref{Picture4_0001}b all the lines are written with the same
$SP=0.2$ V while the tip bias voltage varied between -20 and -18 V for the
lines marked by dashed arrows (left to right) and -16 V for the rightmost line
marked with the solid arrow. This rightmost line is a bump rather than a
trench. All the lines have similar morphology and are indistinguishable in a
frictional image (Fig.~\ref{Picture4_0001}c). We speculate that at low bias
voltages and low set point voltages, the AFM tip partially oxidize the graphene
into \textit{nonvolatile} graphene oxide (GO) with some oxygen incorporated
into the graphene lattice. GO is known to have a larger layer thickness
\cite{navarro07} than graphene, therefore corresponding to bumps on graphene or
graphite surfaces.

To summarize, we have demonstrated AFM-based LAO on graphene.
The lithography is capable of producing small features ($<25$ nm) with sub-nm
spacial resolution, allowing {\it in situ} monitoring of the device parameters
(such as dimensions or electrical conduction) during the fabrication and
easy tuning of the fabrication. We also found that we can write either
trenches or bumps on the graphene surface depending on the lithography
conditions. We attribute bumps to partial oxidation of the graphene with oxygen
incorporated into the graphene lattice.

The work was partially supported by NSF Grant No. ECS-0348289. YPC
gratefully acknowledge the support by the Nanoelectronics Research Initiative (NRI) through the
Midwest Institute of Nanoelectronics Discovery (MIND) and by the Miller Family endowment.


\begin{thebibliography}{23}
\expandafter\ifx\csname natexlab\endcsname\relax\def\natexlab#1{#1}\fi
\expandafter\ifx\csname bibnamefont\endcsname\relax
  \def\bibnamefont#1{#1}\fi
\expandafter\ifx\csname bibfnamefont\endcsname\relax
  \def\bibfnamefont#1{#1}\fi
\expandafter\ifx\csname citenamefont\endcsname\relax
  \def\citenamefont#1{#1}\fi
\expandafter\ifx\csname url\endcsname\relax
  \def\url#1{\texttt{#1}}\fi
\expandafter\ifx\csname urlprefix\endcsname\relax\def\urlprefix{URL }\fi
\providecommand{\bibinfo}[2]{#2}
\providecommand{\eprint}[2][]{\url{#2}}

\bibitem[{\citenamefont{Novoselov et~al.}(2004)\citenamefont{Novoselov, Geim,
  Morozov, Jiang, Zhang, Dubonos, Grigorieva, and Firsov}}]{Novoselov04}
\bibinfo{author}{\bibfnamefont{K.}~\bibnamefont{Novoselov}},
  \bibinfo{author}{\bibfnamefont{A.}~\bibnamefont{Geim}},
  \bibinfo{author}{\bibfnamefont{S.}~\bibnamefont{Morozov}},
  \bibinfo{author}{\bibfnamefont{D.}~\bibnamefont{Jiang}},
  \bibinfo{author}{\bibfnamefont{Y.}~\bibnamefont{Zhang}},
  \bibinfo{author}{\bibfnamefont{S.}~\bibnamefont{Dubonos}},
  \bibinfo{author}{\bibfnamefont{I.}~\bibnamefont{Grigorieva}},
  \bibnamefont{and} \bibinfo{author}{\bibfnamefont{A.}~\bibnamefont{Firsov}},
  \bibinfo{journal}{Science} \textbf{\bibinfo{volume}{306}},
  \bibinfo{pages}{666} (\bibinfo{year}{2004}).

\bibitem[{\citenamefont{Zhang et~al.}(2005)\citenamefont{Zhang, Tan, Stormer,
  and Kim}}]{zhang05}
\bibinfo{author}{\bibfnamefont{Y.}~\bibnamefont{Zhang}},
  \bibinfo{author}{\bibfnamefont{Y.-W.} \bibnamefont{Tan}},
  \bibinfo{author}{\bibfnamefont{H.~L.} \bibnamefont{Stormer}},
  \bibnamefont{and} \bibinfo{author}{\bibfnamefont{P.}~\bibnamefont{Kim}},
  \bibinfo{journal}{Nature} \textbf{\bibinfo{volume}{438}},
  \bibinfo{pages}{201} (\bibinfo{year}{2005}).

\bibitem[{\citenamefont{Novoselov et~al.}(2005)\citenamefont{Novoselov, Geim,
  Morozov, Jiang, Katsnelson, Grigorieva, Dubonos, and Firsov}}]{Novoselov05}
\bibinfo{author}{\bibfnamefont{K.}~\bibnamefont{Novoselov}},
  \bibinfo{author}{\bibfnamefont{A.}~\bibnamefont{Geim}},
  \bibinfo{author}{\bibfnamefont{S.}~\bibnamefont{Morozov}},
  \bibinfo{author}{\bibfnamefont{D.}~\bibnamefont{Jiang}},
  \bibinfo{author}{\bibfnamefont{M.}~\bibnamefont{Katsnelson}},
  \bibinfo{author}{\bibfnamefont{I.}~\bibnamefont{Grigorieva}},
  \bibinfo{author}{\bibfnamefont{S.}~\bibnamefont{Dubonos}}, \bibnamefont{and}
  \bibinfo{author}{\bibfnamefont{A.}~\bibnamefont{Firsov}},
  \bibinfo{journal}{Nature} \textbf{\bibinfo{volume}{438}},
  \bibinfo{pages}{197} (\bibinfo{year}{2005}).

\bibitem[{\citenamefont{Berger et~al.}(2006)\citenamefont{Berger, Song, Li, Wu,
  Brown, Naud, Mayou, Li, Hass, Marchenkov et~al.}}]{Berger06}
\bibinfo{author}{\bibfnamefont{C.}~\bibnamefont{Berger}},
  \bibinfo{author}{\bibfnamefont{Z.}~\bibnamefont{Song}},
  \bibinfo{author}{\bibfnamefont{X.}~\bibnamefont{Li}},
  \bibinfo{author}{\bibfnamefont{X.}~\bibnamefont{Wu}},
  \bibinfo{author}{\bibfnamefont{N.}~\bibnamefont{Brown}},
  \bibinfo{author}{\bibfnamefont{C.}~\bibnamefont{Naud}},
  \bibinfo{author}{\bibfnamefont{D.}~\bibnamefont{Mayou}},
  \bibinfo{author}{\bibfnamefont{T.}~\bibnamefont{Li}},
  \bibinfo{author}{\bibfnamefont{J.}~\bibnamefont{Hass}},
  \bibinfo{author}{\bibfnamefont{A.~N.} \bibnamefont{Marchenkov}},
  \bibnamefont{et~al.}, \bibinfo{journal}{Science}
  \textbf{\bibinfo{volume}{312}}, \bibinfo{pages}{1191} (\bibinfo{year}{2006}).

\bibitem[{\citenamefont{Novoselov et~al.}(2007)\citenamefont{Novoselov, Jiang,
  Zhang, Morozov, Stormer, Zeitler, Maan, Boebinger, Kim, and
  Geim}}]{Novoselov07}
\bibinfo{author}{\bibfnamefont{K.}~\bibnamefont{Novoselov}},
  \bibinfo{author}{\bibfnamefont{Z.}~\bibnamefont{Jiang}},
  \bibinfo{author}{\bibfnamefont{Y.}~\bibnamefont{Zhang}},
  \bibinfo{author}{\bibfnamefont{S.}~\bibnamefont{Morozov}},
  \bibinfo{author}{\bibfnamefont{H.~L.} \bibnamefont{Stormer}},
  \bibinfo{author}{\bibfnamefont{U.}~\bibnamefont{Zeitler}},
  \bibinfo{author}{\bibfnamefont{J.}~\bibnamefont{Maan}},
  \bibinfo{author}{\bibfnamefont{G.}~\bibnamefont{Boebinger}},
  \bibinfo{author}{\bibfnamefont{P.}~\bibnamefont{Kim}}, \bibnamefont{and}
  \bibinfo{author}{\bibfnamefont{A.}~\bibnamefont{Geim}},
  \bibinfo{journal}{Science} \textbf{\bibinfo{volume}{315}},
  \bibinfo{pages}{1379} (\bibinfo{year}{2007}).

\bibitem[{\citenamefont{Geim and Novoselov}(2007)}]{Geim07}
\bibinfo{author}{\bibfnamefont{A.~K.} \bibnamefont{Geim}} \bibnamefont{and}
  \bibinfo{author}{\bibfnamefont{K.~S.} \bibnamefont{Novoselov}},
  \bibinfo{journal}{Nature Mater.} \textbf{\bibinfo{volume}{6}},
  \bibinfo{pages}{183} (\bibinfo{year}{2007}).

\bibitem[{\citenamefont{Russo et~al.}(2008)\citenamefont{Russo, Oostinga,
  Wehenkel, Heersche, Sobhani, Vandersypen, and Morpurgo}}]{Russo08}
\bibinfo{author}{\bibfnamefont{S.}~\bibnamefont{Russo}},
  \bibinfo{author}{\bibfnamefont{J.~B.} \bibnamefont{Oostinga}},
  \bibinfo{author}{\bibfnamefont{D.}~\bibnamefont{Wehenkel}},
  \bibinfo{author}{\bibfnamefont{H.~B.} \bibnamefont{Heersche}},
  \bibinfo{author}{\bibfnamefont{S.~S.} \bibnamefont{Sobhani}},
  \bibinfo{author}{\bibfnamefont{L.~M.~K.} \bibnamefont{Vandersypen}},
  \bibnamefont{and} \bibinfo{author}{\bibfnamefont{A.~F.}
  \bibnamefont{Morpurgo}}, \bibinfo{journal}{Phys. Rev. B}
  \textbf{\bibinfo{volume}{77}}, \bibinfo{pages}{085413}
  (\bibinfo{year}{2008}).

\bibitem[{\citenamefont{Han et~al.}(2007)\citenamefont{Han, Oezyilmaz, Zhang,
  and Kim}}]{Han07}
\bibinfo{author}{\bibfnamefont{M.~Y.} \bibnamefont{Han}},
  \bibinfo{author}{\bibfnamefont{B.}~\bibnamefont{Oezyilmaz}},
  \bibinfo{author}{\bibfnamefont{Y.}~\bibnamefont{Zhang}}, \bibnamefont{and}
  \bibinfo{author}{\bibfnamefont{P.}~\bibnamefont{Kim}},
  \bibinfo{journal}{Phys. Rev. Lett.} \textbf{\bibinfo{volume}{98}},
  \bibinfo{pages}{206805} (\bibinfo{year}{2007}).

\bibitem[{\citenamefont{Chen et~al.}(2007)\citenamefont{Chen, Lin, Rooks, and
  Avouris}}]{Chen07}
\bibinfo{author}{\bibfnamefont{Z.}~\bibnamefont{Chen}},
  \bibinfo{author}{\bibfnamefont{Y.-M.} \bibnamefont{Lin}},
  \bibinfo{author}{\bibfnamefont{M.~J.} \bibnamefont{Rooks}}, \bibnamefont{and}
  \bibinfo{author}{\bibfnamefont{P.}~\bibnamefont{Avouris}},
  \bibinfo{journal}{Phys. E} \textbf{\bibinfo{volume}{40}},
  \bibinfo{pages}{228} (\bibinfo{year}{2007}).

\bibitem[{\citenamefont{Bunch et~al.}(2005)\citenamefont{Bunch, Yaish, Brink,
  Bolotin, and McEuen}}]{Bunch05}
\bibinfo{author}{\bibfnamefont{J.~S.} \bibnamefont{Bunch}},
  \bibinfo{author}{\bibfnamefont{Y.}~\bibnamefont{Yaish}},
  \bibinfo{author}{\bibfnamefont{M.}~\bibnamefont{Brink}},
  \bibinfo{author}{\bibfnamefont{K.}~\bibnamefont{Bolotin}}, \bibnamefont{and}
  \bibinfo{author}{\bibfnamefont{P.~L.} \bibnamefont{McEuen}},
  \bibinfo{journal}{Nano Lett.} \textbf{\bibinfo{volume}{5}},
  \bibinfo{pages}{287} (\bibinfo{year}{2005}).

\bibitem[{\citenamefont{Stampfer et~al.}(2008)\citenamefont{Stampfer,
  Guettinger, Molitor, Graf, Ihn, and Ensslin}}]{Stampfer08}
\bibinfo{author}{\bibfnamefont{C.}~\bibnamefont{Stampfer}},
  \bibinfo{author}{\bibfnamefont{J.}~\bibnamefont{Guettinger}},
  \bibinfo{author}{\bibfnamefont{F.}~\bibnamefont{Molitor}},
  \bibinfo{author}{\bibfnamefont{D.}~\bibnamefont{Graf}},
  \bibinfo{author}{\bibfnamefont{T.}~\bibnamefont{Ihn}}, \bibnamefont{and}
  \bibinfo{author}{\bibfnamefont{K.}~\bibnamefont{Ensslin}},
  \bibinfo{journal}{Appl. Phys. Lett.} \textbf{\bibinfo{volume}{92}},
  \bibinfo{pages}{012102} (\bibinfo{year}{2008}).

\bibitem[{\citenamefont{Tseng et~al.}(2005)\citenamefont{Tseng, Notargiacomo,
  and Chen}}]{tseng05}
\bibinfo{author}{\bibfnamefont{A.~A.} \bibnamefont{Tseng}},
  \bibinfo{author}{\bibfnamefont{A.}~\bibnamefont{Notargiacomo}},
  \bibnamefont{and} \bibinfo{author}{\bibfnamefont{T.~P.} \bibnamefont{Chen}},
  \bibinfo{journal}{J Vac. Sci. Tech. B} \textbf{\bibinfo{volume}{23}},
  \bibinfo{pages}{877} (\bibinfo{year}{2005}).

\bibitem[{\citenamefont{Snow and Campbell}(1994)}]{snow94}
\bibinfo{author}{\bibfnamefont{E.~S.} \bibnamefont{Snow}} \bibnamefont{and}
  \bibinfo{author}{\bibfnamefont{P.~M.} \bibnamefont{Campbell}},
  \bibinfo{journal}{\apl} \textbf{\bibinfo{volume}{64}}, \bibinfo{pages}{1932}
  (\bibinfo{year}{1994}).

\bibitem[{\citenamefont{Held et~al.}(1997)\citenamefont{Held, Heinzel,
  Studerus, Ensslin, and Holland}}]{held97}
\bibinfo{author}{\bibfnamefont{R.}~\bibnamefont{Held}},
  \bibinfo{author}{\bibfnamefont{T.}~\bibnamefont{Heinzel}},
  \bibinfo{author}{\bibfnamefont{A.~P.} \bibnamefont{Studerus}},
  \bibinfo{author}{\bibfnamefont{K.}~\bibnamefont{Ensslin}}, \bibnamefont{and}
  \bibinfo{author}{\bibfnamefont{M.}~\bibnamefont{Holland}},
  \bibinfo{journal}{\apl} \textbf{\bibinfo{volume}{71}}, \bibinfo{pages}{2689}
  (\bibinfo{year}{1997}).

\bibitem[{\citenamefont{Rokhinson et~al.}(2002)\citenamefont{Rokhinson, Tsui,
  Pfeiffer, and West}}]{rokhinson02}
\bibinfo{author}{\bibfnamefont{L.~P.} \bibnamefont{Rokhinson}},
  \bibinfo{author}{\bibfnamefont{D.~C.} \bibnamefont{Tsui}},
  \bibinfo{author}{\bibfnamefont{L.~N.} \bibnamefont{Pfeiffer}},
  \bibnamefont{and} \bibinfo{author}{\bibfnamefont{K.~W.} \bibnamefont{West}},
  \bibinfo{journal}{Superlattices Microstruct.} \textbf{\bibinfo{volume}{32}},
  \bibinfo{pages}{99} (\bibinfo{year}{2002}).

\bibitem[{\citenamefont{Kim et~al.}(2003)\citenamefont{Kim, Koo, and
  Kim}}]{kim03}
\bibinfo{author}{\bibfnamefont{D.~H.} \bibnamefont{Kim}},
  \bibinfo{author}{\bibfnamefont{J.~Y.} \bibnamefont{Koo}}, \bibnamefont{and}
  \bibinfo{author}{\bibfnamefont{J.~J.} \bibnamefont{Kim}},
  \bibinfo{journal}{\prb} \textbf{\bibinfo{volume}{68}},
  \bibinfo{pages}{113406} (\bibinfo{year}{2003}).

\bibitem[{\citenamefont{Park et~al.}(2007)\citenamefont{Park, Zhang, Liang, and
  Wang}}]{park07}
\bibinfo{author}{\bibfnamefont{J.~G.} \bibnamefont{Park}},
  \bibinfo{author}{\bibfnamefont{C.}~\bibnamefont{Zhang}},
  \bibinfo{author}{\bibfnamefont{R.}~\bibnamefont{Liang}}, \bibnamefont{and}
  \bibinfo{author}{\bibfnamefont{B.}~\bibnamefont{Wang}},
  \bibinfo{journal}{Nanotechnology} \textbf{\bibinfo{volume}{18}},
  \bibinfo{pages}{405306} (\bibinfo{year}{2007}).

\bibitem[{\citenamefont{Staley et~al.}(2008)\citenamefont{Staley, Puls, and
  Liu}}]{staley08}
\bibinfo{author}{\bibfnamefont{N.}~\bibnamefont{Staley}},
  \bibinfo{author}{\bibfnamefont{C.}~\bibnamefont{Puls}}, \bibnamefont{and}
  \bibinfo{author}{\bibfnamefont{Y.}~\bibnamefont{Liu}},
  \bibinfo{journal}{\prb} \textbf{\bibinfo{volume}{77}},
  \bibinfo{pages}{155429} (\bibinfo{year}{2008}).

\bibitem[{\citenamefont{Graf et~al.}(2007)\citenamefont{Graf, Molitor, T.Ihn,
  and Ensslin}}]{graf07}
\bibinfo{author}{\bibfnamefont{D.}~\bibnamefont{Graf}},
  \bibinfo{author}{\bibfnamefont{F.}~\bibnamefont{Molitor}},
  \bibinfo{author}{\bibnamefont{T.Ihn}}, \bibnamefont{and}
  \bibinfo{author}{\bibfnamefont{K.}~\bibnamefont{Ensslin}},
  \bibinfo{journal}{\prb} \textbf{\bibinfo{volume}{75}},
  \bibinfo{pages}{245429} (\bibinfo{year}{2007}).

\bibitem[{\citenamefont{Moriki et~al.}(2007)\citenamefont{Moriki, Kanda, Sato,
  Miyazaki, Odaka, Ootuka, Aoyagi, and Tsukagoshi}}]{moriki07}
\bibinfo{author}{\bibfnamefont{T.}~\bibnamefont{Moriki}},
  \bibinfo{author}{\bibfnamefont{A.}~\bibnamefont{Kanda}},
  \bibinfo{author}{\bibfnamefont{T.}~\bibnamefont{Sato}},
  \bibinfo{author}{\bibfnamefont{H.}~\bibnamefont{Miyazaki}},
  \bibinfo{author}{\bibfnamefont{S.}~\bibnamefont{Odaka}},
  \bibinfo{author}{\bibfnamefont{Y.}~\bibnamefont{Ootuka}},
  \bibinfo{author}{\bibfnamefont{Y.}~\bibnamefont{Aoyagi}}, \bibnamefont{and}
  \bibinfo{author}{\bibfnamefont{K.}~\bibnamefont{Tsukagoshi}},
  \bibinfo{journal}{Phys. E} \textbf{\bibinfo{volume}{40}},
  \bibinfo{pages}{241} (\bibinfo{year}{2007}).

\bibitem[{\citenamefont{Recher et~al.}(2007)\citenamefont{Recher, Trauzettel,
  Rycerz, Blanter, Beenakker, and Morpurgo}}]{recher07}
\bibinfo{author}{\bibfnamefont{P.}~\bibnamefont{Recher}},
  \bibinfo{author}{\bibfnamefont{B.}~\bibnamefont{Trauzettel}},
  \bibinfo{author}{\bibfnamefont{A.}~\bibnamefont{Rycerz}},
  \bibinfo{author}{\bibfnamefont{Y.~M.} \bibnamefont{Blanter}},
  \bibinfo{author}{\bibfnamefont{C.~W.~J.} \bibnamefont{Beenakker}},
  \bibnamefont{and} \bibinfo{author}{\bibfnamefont{A.~F.}
  \bibnamefont{Morpurgo}}, \bibinfo{journal}{\prb}
  \textbf{\bibinfo{volume}{76}}, \bibinfo{pages}{235404}
  (\bibinfo{year}{2007}).

\bibitem[{\citenamefont{Kondo et~al.}(1995)\citenamefont{Kondo, Heike,
  Lutwyche, and Wada}}]{kondo95}
\bibinfo{author}{\bibfnamefont{S.}~\bibnamefont{Kondo}},
  \bibinfo{author}{\bibfnamefont{S.}~\bibnamefont{Heike}},
  \bibinfo{author}{\bibfnamefont{M.}~\bibnamefont{Lutwyche}}, \bibnamefont{and}
  \bibinfo{author}{\bibfnamefont{Y.}~\bibnamefont{Wada}}, \bibinfo{journal}{J.
  Appl. Phys.} \textbf{\bibinfo{volume}{78}}, \bibinfo{pages}{155}
  (\bibinfo{year}{1995}).

\bibitem[{\citenamefont{G\'{o}mez-Navarro
  et~al.}(2007)\citenamefont{G\'{o}mez-Navarro, Weitz, Bittner, Scolari, Mews,
  Burghard, and Kern}}]{navarro07}
\bibinfo{author}{\bibfnamefont{C.}~\bibnamefont{G\'{o}mez-Navarro}},
  \bibinfo{author}{\bibfnamefont{R.~T.} \bibnamefont{Weitz}},
  \bibinfo{author}{\bibfnamefont{A.~M.} \bibnamefont{Bittner}},
  \bibinfo{author}{\bibfnamefont{M.}~\bibnamefont{Scolari}},
  \bibinfo{author}{\bibfnamefont{A.}~\bibnamefont{Mews}},
  \bibinfo{author}{\bibfnamefont{M.}~\bibnamefont{Burghard}}, \bibnamefont{and}
  \bibinfo{author}{\bibfnamefont{K.}~\bibnamefont{Kern}},
  \bibinfo{journal}{Nano Lett.} \textbf{\bibinfo{volume}{7}},
  \bibinfo{pages}{3499} (\bibinfo{year}{2007}).

\end{thebibliography}

\end{document}